\documentclass[twocolumn,preprintnumbers,amsmath,amssymb]{revtex4}

\usepackage{graphicx}
\usepackage{dcolumn}
\usepackage{bm}
\usepackage{rotating}
\usepackage{verbatim}
\begin{document}

\title{Model for Incomplete Reconnection in Sawtooth Crashes}

\author{M.~T.~Beidler and P.~A.~Cassak} 
\affiliation{Department of Physics,
West Virginia University, Morgantown, WV, 26506, USA}

\preprint{Accepted to Phys.~Rev.~Lett., November 1, 2011}

\begin{abstract}
  A model for incomplete reconnection in sawtooth crashes is
  presented. The reconnection inflow during the crash phase of sawteeth
  self-consistently convects the high pressure core toward the
  reconnection site, raising the pressure gradient there.
  Reconnection shuts off if the diamagnetic drift speed at the
  reconnection site exceeds a threshold, which may explain incomplete
  reconnection.  The relaxation of magnetic shear after reconnection
  stops may explain the destabilization of ideal interchange
  instabilities reported previously. Proof-of-principle two-fluid
  simulations confirm this basic picture.  Predictions of the model
  compare favorably to data from the Mega Ampere Spherical Tokamak.
  Applications to transport modeling of sawteeth are discussed. The
  results should apply across tokamaks, including ITER.

\end{abstract}

\maketitle


Sawtooth crashes in tokamaks occur when the core temperature rapidly
drops following a slow rise \cite{vonGoeler74}. Large sawteeth are
deleterious for fusion because they spoil confinement, while
small sawteeth may be beneficial by limiting impurity accumulation
\cite{Hender07}. Kadomtsev suggested the cause is the $m=1,n=1$
tearing mode \cite{Kadomtsev75b}, where $m$ and $n$ are poloidal and
toroidal mode numbers.  The predicted crash duration is the time it
takes Sweet-Parker reconnection to process all available magnetic
flux. This agreed with early experiments and simulations.

Soon after, cracks in the model appeared.  Crash times in larger and
hotter tokamaks were much faster than Kadomtsev's prediction
\cite{Edwards86,Yamada94}. Also, Kadomtsev's model assumes all
available magnetic flux reconnects (reconnection is ``complete''),
however experiments reveal that reconnection is usually incomplete
\cite{Soltwisch92}. Equivalently, the safety factor $q = r B_{\varphi}
/ R_0 B_{\theta}$ does not exceed 1 everywhere after a crash, where
$R_0$ and $r$ are the major and minor radii and $B_{\varphi}$ and
$B_{\theta}$ are toroidal and poloidal magnetic fields.

Many models of incomplete reconnection exist, but there is no
consensus on which, if any, is correct.  Examples include stochastic
magnetic fields \cite{Lichtenberg92}, diamagnetic and pressure effects
at the magnetic island \cite{Biskamp81,Biskamp97b,Park87,Wang95},
trapped high energy particles \cite{Coppi88,White89,Porcelli91a}, a
flattened $q$-profile \cite{Holmes89}, and the presence of shear flow
\cite{Kleva92,Kleva02}.

The uncertainty of the cause of incomplete reconnection impacts
tokamak transport modeling. Low-dimensional transport models capture
the sawtooth period and amplitude \cite{Porcelli96}, but the fraction
of flux reconnected is an input parameter rather than
self-consistently calculated. A self-consistent theory of incomplete
reconnection would improve tokamak transport models.


In this letter, we propose a model for incomplete reconnection in
sawteeth due to the self-consistent dynamics of magnetic reconnection,
building on established properties of diamagnetic effects
\cite{Swisdak03}.  After describing the model, we present numerical
simulations confirming its key aspects. Then, we show that the model
is consistent with data from the Mega Ampere Spherical Tokamak (MAST)
\cite{Chapman10}.  Finally, applications and limitations of the result
are discussed.

\begin{figure}
\includegraphics[width=3.4in]{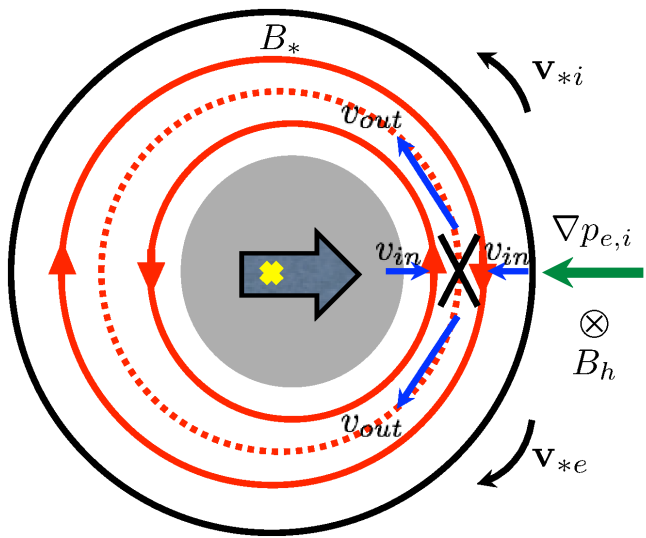}
\caption{\label{fig-kink} (Color) Sketch of the $m=1,n=1$
  reconnection plane. Reconnecting (auxiliary) magnetic fields $B_*$
  are in red with the rational surface $r_s$ indicated by the dotted
  red line. Plasma inflows $v_{in}$ and outflows $v_{out}$ are in blue
  with the reconnection site at the black X. The grey core moves from
  its initial position centered at the yellow X. The pressure gradient
  is the green arrow. The helical guide field $B_h$ and the
  diamagnetic drift velocities ${\bf v}_{*i}$ and ${\bf v}_{*e}$ are
  shown.}
\end{figure}

To understand why reconnection in Kadomtsev's model is complete,
consider the $m=1,n=1$ reconnection plane sketched in
Fig.~\ref{fig-kink}. The reversed (auxiliary) magnetic field $B_*$ is
in red, the high pressure core is in grey, and the reconnection site
is the black X.  When reconnection begins, outflow jets (in blue) are
driven by tension in newly reconnected field lines. Mass continuity
induces plasma inflow from upstream (also in blue). This flow convects
more magnetic flux (if available) towards the reconnection site, which
reconnects. Thus, reconnection is self-sustaining.

We argue that the key to explaining incomplete reconnection is the
effect of reconnection dynamics on the pressure gradient at the
reconnection site. Suppose the core is initially centered at the
yellow X. The pressure gradient at the reconnection site (the green
arrow) is radially inward and relatively weak. As the reconnection
inflow self-consistently convects the core outward, the pressure
gradient at the reconnection site increases. The outward motion of the
core has long been seen in observations \cite{Yamada94}.

In the presence of a strong out-of-plane (guide) magnetic field $B_h$,
in-plane pressure gradients lead to in-plane diamagnetic drifts,
sketched in Fig.~\ref{fig-kink}.  Diamagnetic ($\omega_{*}$) effects
are known to stabilize linear and nonlinear tearing
\cite{Zakharov93,Rogers95}, which continues to be actively studied
\cite{Swisdak03,Germaschewski06,Bhattacharjee08}. It was shown
\cite{Swisdak03} that reconnection does not occur if
\begin{equation}
  |{\bf v}_{*i} - {\bf v}_{*e}|_{out} > v_{out},  
\end{equation}
where $v_{out}$ is the reconnection outflow speed, ${\bf v}_{* \alpha}
= - {\bf \nabla}p_{\alpha} \times {\bf B}/(q_{\alpha}n_{\alpha}B^2)$
is the diamagnetic drift velocity measured at the reconnection site
for species $\alpha = i,e$, and the ``out'' subscript refers to the
outflow direction.

We propose that the increase in $v_{*i}$ and $v_{*e}$ as the pressure
gradient self-consistently increases due to reconnection causes the
left-hand side of Eq.~(1) to increase. If Eq.~(1) is never satisfied,
reconnection is complete. If the pressure gradient becomes large
enough, reconnection ceases. Since Eq.~(1) can be satisfied even when
free magnetic energy remains, this provides a possible mechanism for
incomplete reconnection.  This model departs from previous ones
\cite{Biskamp81,Biskamp97b,Park87} as it concerns pressure gradients
at the reconnection site rather than the magnetic islands.
 
This model complements, and may explain key global features of, recent
observations at MAST \cite{Chapman10}. They observe that $|\nabla
T_e|$ increases during a sawtooth period, peaking at the end of the
crash (their Fig.~3), qualitatively consistent with the model. They
also show that secondary ideal-MHD instabilities are destabilized at
the end of the crash cycle. Reconnection would also play an important
role in this process. When reconnection ceases, the electron-scale
current sheet broadens, reducing the magnetic shear in a region where
$|\nabla p|$ is large. Decreased shear is known to destabilize
interchange instabilities ({\it e.g.} \cite{Freidberg87}).

 
To test the model, proof-of-principle numerical simulations are
performed using F3D \cite{Shay04}, a two-fluid code employing a
two-dimensional slab geometry with periodic boundary conditions. This
geometry is appropriate because motion in the plane normal to the
guide magnetic field is well described in two dimensions, toroidal
effects are not expected to play a role on the short time scales in
question (tens of $\mu {\rm s}$), and three-dimensional toroidal
simulations employ unphysical forcing terms to obtain sawteeth
\cite{Breslau08}. These simulations do not contain toroidal effects
which lead to secondary ideal-MHD instabilities \cite{Chapman10}
because this facet of the evolution is outside the scope of this
study. Electron pressure is evolved assuming an adiabatic ideal gas
with a ratio of electron specific heats $\gamma_e = 5/3$. Since the
relative diamagnetic speed is the key parameter, ions are assumed cold
for simplicity. Magnetic fields and mass densities are normalized to
arbitrary values $B_{0}$ and $\rho_{0}$, velocities to the Alfv\'en
speed $c_{A0} = B_{0} / (4 \pi \rho_{0})^{1/2}$, lengths to the ion
inertial length $d_{i0} = c / \omega_{pi} = (m_i^2c^2/4\pi \rho_0
Z_{{\rm eff}}^2 e^2)^{1/2}$, times to the ion cyclotron time
$\Omega_{ci0}^{-1} = (Z_{{\rm eff}}eB_0/m_ic)^{-1}$, electric fields
to $E_{0} = c_{A0} B_{0} /c$, and pressures to $p_0 = B_0^2/4\pi$,
where $m_{i}$ is the ion mass, $c$ is the speed of light, $e$ is the
proton charge, and $Z_{{\rm eff}}$ is the effective atomic number.

The coordinate system has $x$ parallel to the inflow (radial), $y$
parallel to the outflow (poloidal), and $z$ in the out-of-plane
(toroidal) direction, invariant in the present two-dimensional
simulations. The equilibrium has an in-plane magnetic field profile of
a double Harris sheet,
\[
B_{y}(x) = \tanh\left(\frac{x-L_{x}/4}{w_{0}}\right) -
\tanh\left(\frac{x+L_{x}/4}{w_{0}}\right) + 1,
\]
where $L_{x} \times L_{y} = 102.4 \times 204.8$ is the system size and
$w_{0} = 0.5$ is the initial thickness of the current sheet.  For this
equilibrium, the toroidal mode number $n = 0$ manifestly, so the
rational surfaces are $x_{s} = \pm L_{x} / 4 = \pm 25.6$. We focus on
a single mode because there is typically a dominant mode in sawteeth;
the $n=0$ mode is chosen for simplicity, but is not expected to
alter the conclusions. The mass density is initially $\rho = 1$.  The
initial electron pressure profile is
\begin{eqnarray}
  p_{e}(x) \! \! \! \!  && =   \frac{1}{2} \left( p_{1} + p_{2} \right) + \frac{1}{2}
  \left( p_{1} - p_{2} \right) \times \nonumber \\ && \left[ \tanh\left(\frac{x+3
        L_{x}/8}{w_{p}}\right) - \tanh\left(\frac{x-3
        L_{x}/8}{w_{p}}\right) - 1 \right]. \nonumber 
\end{eqnarray}
The pressure gradient is localized near $x = \pm 3 L_{x} / 8 = \pm
38.4$ rather than at the rational surfaces $x_{s}$.  Thus, $p_{e}$ at
the reconnection site is initially uniform. The length scale of the
pressure gradient is $w_{p}$ = 2. The guide magnetic field $B_{z}(x)$
has a mean value of $5$ with a profile that ensures initial pressure
balance, $p + B^{2} / 2  = {\rm constant}$.

The data we present are from simulations with a grid scale of $\Delta =
0.05$. A test simulation with $\Delta = 0.025$ confirms the resolution
is sufficient.  The equations employ fourth-order diffusion with
coefficient $D_{4} = 2 \times 10^{-5}$ to damp noise at the grid
scale; $D_4$ has been varied to ensure the key physics is not
sensitive to it. The electron to ion mass ratio is 1/25. Simulations
include no resistivity because experimental crash times are faster
than collisional reconnection times. The presented simulations do not
employ a parallel thermal conductivity, but test simulations with
$\chi_{||} = 0.08$ reveal no significant changes. Tearing is initiated
by a small coherent perturbation to the in-plane magnetic field of
amplitude $0.01$. It is known that secondary islands can spontaneously
arise in reconnection simulations; due to symmetry, such islands would
stay at the original X-line \cite{Loureiro05}. To prevent this,
initial random magnetic perturbations of magnitude $2.0 \times
10^{-5}$ break symmetry so secondary islands are ejected.

The principal simulation employs $p_{1} = 5, p_{2} = 25$ so $v_{*e}$
will exceed $v_{out}$ when the high pressure plasma convects in. Other
simulation parameters are carefully chosen: $B_{z} \gg B_y$ as is
relevant to sawteeth and $p_{e}$ is large enough so the ion Larmor
radius $\rho_{s} = c_{s} / \Omega_{ci}$ exceeds the electron skin
depth $d_{e}=c/\omega_{pe}$, allowing fast reconnection to proceed
\cite{Aydemir92,Rogers01}. Here, $c_{s} = (\gamma_{e} Z_{{\rm eff}}
T_{e} / m_{i})^{1/2}$ is the ion acoustic speed, and $T_{e}$ is the
electron temperature.

\begin{figure}
\includegraphics[width=3.4in]{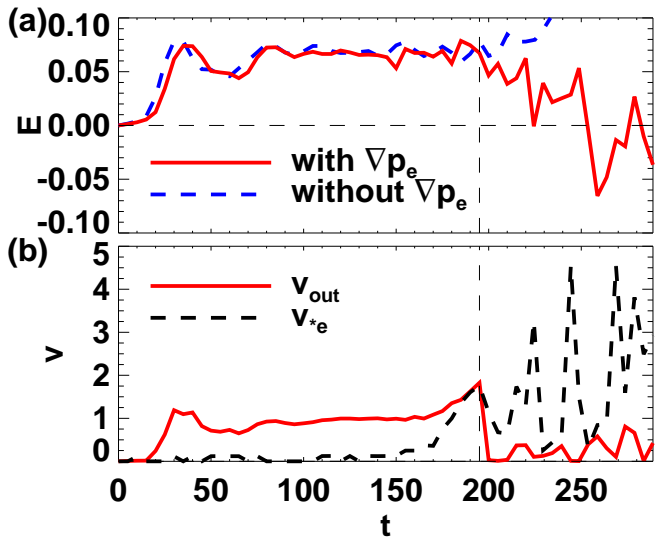}
\caption{\label{fig-rrate} (Color online) (a) Reconnection rate $E$ as
  a function of time $t$ with and without a pressure gradient. (b)
  Diamagnetic drift speed $v_{*e}$ at the reconnection site and
  outflow speed $v_{out}$ vs.~$t$.}
\end{figure}

Upon evolving the system, Hall reconnection occurs initially and the
high pressure plasma convects towards the reconnection site as
expected. The reconnection rate $E$, measured as the time rate of
change of magnetic flux between the X-line and O-line, is plotted as
the solid (red) line in Fig.~\ref{fig-rrate}(a).  It increases from
zero to its expected value near 0.1 \cite{Shay99} by $t \sim 90$,
where it reaches a steady-state with a single X-line. (The variation
between $t = 40$ and 90 is due to transient secondary island formation
and coalescence.)  At $t \simeq 195$, $E$ begins decreasing. It
decreases to below zero, where it fluctuates for a number of
Alfv\'en crossing times. Thus, reconnection has shut off.

To determine the cause, the electron diamagnetic speed $v_{*e}$ at the
reconnection site is plotted as a function of time in
Fig.~\ref{fig-rrate}(b) as the dashed (black) line. For comparison,
the outflow speed $v_{out}$ is plotted as the solid (red)
line. Asymmetric outflows occur when there is a pressure gradient in
the outflow direction \cite{Murphy10}, and since such gradients
self-consistently generate here, $v_{out}$ is calculated as the
average of the maximum electron outflow speeds from either side of the
reconnection site, averaged over $5 d_e$ when turbulent.

Figure 2(b) reveals that $v_{*e}$ is small initially, but increases in
time once the pressure gradient reaches the reconnection site at $t
\simeq 140$. It increases until it becomes comparable to $v_{out}$ at
$t \simeq 195$ (the vertical dashed line), the same time $E$ begins to
decrease.  Therefore, reconnection is throttled when Eq.~(1) is first
satisfied.

\begin{figure}
\includegraphics[width=3.4in]{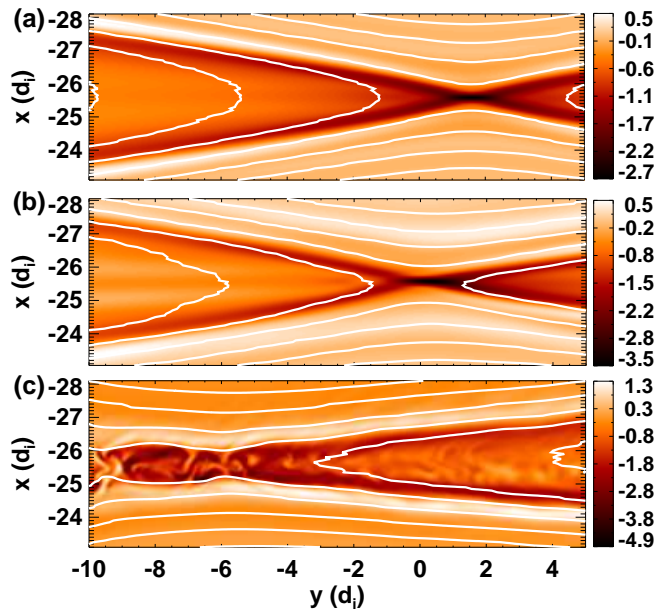}
\caption{\label{fig-curz} (Color) Out-of-plane current density $J_z$
  zoomed in near the X-line with magnetic field lines superimposed (a)
  before ($t = 125$), (b) after ($t=180$), and (c) significantly after
  ($t=210$) the pressure gradient reaches the reconnection site. The
  $x$ and $y$ axes correspond to the radial and poloidal directions,
  respectively.}
\end{figure}

To ensure diamagnetic effects occur, the out-of-plane current density
$J_z$ near the X-line is plotted in Fig.~\ref{fig-curz} (a) before ($t
= 125$) and (b) after ($t = 180$) the pressure gradient arrives, with
in-plane magnetic field lines superimposed. The guide field is in the
$-z$-direction and $\nabla p_e$ is in the $-x$-direction. The
reconnection site drifts in the $-y$-direction, the direction of ${\bf
  v}_{*e}$. Note, a secondary instability (recently speculated to be a
drift instability \cite{Drake10}) appears. The increased variability
of $v_{*e}$ and $E$ after $t \simeq 205$ are attributed to this
instability.

To ensure the observed effect is caused by the pressure gradient,
simulations with other pressure profiles are performed.  When there is
no gradient with $p_{1} = p_{2} = 5$, there is no decrease in $E$,
plotted as the dashed (blue) line in Fig.~\ref{fig-rrate}(a). The same
is true for $p_{1} = p_{2} = 25$ (not plotted).  When $p_{1} =5, p_{2}
=7$, no drop in reconnection rate is observed because the maximum
$v_{*e}$ only reaches $\sim 1$, but $v_{out} \sim 2$ so Eq.~(1) is
never satisfied. In summary, the simulations confirm the basic
prediction of the model: reconnection ceases when large enough
pressure gradients self-consistently convect into the reconnection
site despite the presence of free magnetic energy.

Post-cessation features are important for the subsequent
dynamics. Figure 3(c) shows $J_z$ significantly after the pressure
gradient reaches the reconnection site ($t=210$). The current layer
clearly broadens as reconnection stops, reducing the magnetic shear at
the reconnection site, as evidenced by the negative reconnection rate
in Fig.~2(a). The reduced shear would make the system more prone to
interchange instabilities, which were argued to occur in
Ref.~\cite{Chapman10}.


Equation (1) provides a quantitative prediction of the conditions at
the end of sawteeth; we assess it with data from MAST
\cite{Chapman10}. To transform into the plane of reconnection
perpendicular to the $m=1,n=1$ helical direction, the reconnecting
(auxiliary) field $B_{*}$ is related to the toroidal $B_{\varphi}$ and
poloidal $B_{\theta}$ fields by
\begin{equation}
B_{*}(r)=B_{\theta}- \left( \frac{r}{R_0} \right) B_{\varphi}.
\end{equation}
At MAST, $R_0 = 0.85 \ {\rm m}$ \cite{Appel08} while $B_{\varphi}
\simeq 0.4 \ {\rm T}$ and $B_{\theta} \simeq 0.15 \ {\rm T}$
\cite{Chapman10b}.  The rational surface $r_s$ is where $B_* = 0$ in
Eq.~(2), which gives $r_s \simeq 0.32 \ {\rm m}$.  This result agrees
well with Fig.~1(a) of Ref.~\cite{Chapman10}.  The helical guide field
at $r_s$ is $B_h = B_{\varphi} ( 1 + r_{s} / R_{0}) \simeq 0.55 \ {\rm
  T}$.

To test the model, Eq.~(1) must be evaluated at the end of the
sawtooth crash. The outflow speed scales with $ c_{Ae}$, the electron
Alfv\'en speed based on the field $B_{*e}$ upstream of the electron
current layer. Assuming the large guide field limit with $B_h \gg B_*$
in the vicinity of $r_s$, the thickness of the electron current layer
scales as the electron Larmor radius $\rho_{e} = v_{th,e}/\Omega_{ce}$
\cite{Horiuchi97}, where $v_{th,e} = (\gamma_e T_e/m_e)^{1/2}$ is the
electron thermal speed and $\Omega_{ce} = e B/m_ec$ is the electron
cyclotron frequency. Using $T_{e} \simeq 500 \ {\rm eV}$ at $r_s$
\cite{Chapman10} and $\gamma_{e} = 5/3$, we find $\rho_{e} \simeq
0.013 \ {\rm cm}$.  To find $B_{*e}$, we evaluate Eq.~(2) at $r_{s}
\pm 2 \rho_{e}$ \cite{Jemella03}, which gives $B_{*e} \simeq 5.9
\times 10^{-5} \ {\rm T}$, justifying the strong guide field
assumption.  Using this value gives $v_{out} \approx 14.2 \ {\rm
  km/s}$, where $n_{e} \simeq 6 \times 10^{19} \ {\rm m^{-3}}$ is
estimated from Fig.~2 in Ref.~\cite{Chapman10}.

To estimate $v_{*e}$, note $|\nabla p_e|/n_e = |\nabla T_e| +
T_e(|\nabla n_e|/n_e)$. The right-hand side is estimated at the end of
the crash from Figs.~1(e), 2 and 3 of Ref.~\cite{Chapman10} to be
$|\nabla p_e|/n_e \simeq 7400 \ {\rm eV/m}$. Then, the electron
diamagnetic speed is $v_{*e} = |\nabla p_e|/(q n_e B_h) \approx 13.5 \
{\rm km/s}$. Equation (1) includes ion diamagnetic effects, but
complementary ion data is unavailable \cite{Chapman10b}.  Assuming the
ion temperature has a similar profile as the electrons with $T_e >
T_i$, we expect $v_{*e} < |v_{*i}|+|v_{*e}| < 2v_{*e}$. Thus, the two
speeds agree rather well, showing the agreement with the data is also
quantitative.

As a further consistency check, we compare the speed of the core to
the inflow speed. The the core's speed is estimated from Figs.~1(d-f)
of Ref.~\cite{Chapman10} by dividing its displacement ($\simeq 0.08 \
{\rm m}$) by the elapsed time ($\simeq 0.04 \ {\rm ms}$), giving a
speed of $\sim 2 \ {\rm km/s}$. The reconnection inflow speed scales
like $0.1 c_{Ai}$ \cite{Shay04}, where $c_{Ai}$ is the ion Alfv\'en
speed based on the field $B_{*i}$ upstream of the ion current
layer. The ion layer thickness with a large guide field scales like
the ion Larmor radius $\rho_{s}$ \cite{Zakharov93}. Using $Z_{{\rm
    eff}} \sim 1$ \cite{Tournianski05} and $m_i = 2 m_p$ for a
deuterium plasma \cite{Appel08}, we find $\rho_s ∼\sim 0.77 \ {\rm
  cm}$. As in the calculation of $B_{*e}$, we evaluate Eq.~(2) at $r_s
\pm 2\rho_s$, giving $B_{*i} = 6.7 \times 10^{-3} \ {\rm T}$. Then,
$c_{Ai} \approx 13 \ {\rm km/s}$, so the inflow speed is $\simeq 1.3 \
{\rm km/s}$. Thus, the inflow speed is comparable to the speed of the
core, as predicted.

For tokamak applications, Eq.~(1) may be recast in terms of more
familiar quantities. Assuming $v_{out} \sim c_{Ae}$ in Eq.~(1) and
rewriting Eq.~(2) in terms of $q$ and expanding to lowest order in $r$
for a small displacement ($2\rho_e$) from $r_s$, $B_{*e} \simeq
B_{\theta} q^{\prime} 2 \rho_e$, where the prime denotes a radial
derivative. Thus, Eq.~(1) becomes
\begin{equation}
  \frac{1}{e B_h} \left| \frac{p_i^{\prime}}{Z_{{\rm eff}}n_i} + \frac{p_e^{\prime}}{n_e} \right| > \frac{2 \rho_e B_{\theta}}{\sqrt{4 \pi m_e n_e}} q^{\prime}, 
\end{equation}
where all quantities are evaluated at $r_s$. This expression is
reminiscent of the condition on $p^{\prime}$ and $q^{\prime}$ for
suppression of sawteeth derived from linear tearing theory
\cite{Zakharov93,Levinton94}.


In conclusion, we have described a model for incomplete reconnection
in sawtooth crashes, tested the basic physics with numerical
simulations, and shown it is consistent with data from MAST.
Interestingly, recent simulations of sawteeth revealed complete
reconnection in MHD, but incomplete reconnection in extended-MHD with
electron and ion diamagnetic effects \cite{Breslau07,Breslau08}; the
present result may be relevant. Equation (1) may be useful for
low-dimensional transport modeling, which currently use ad hoc models
to achieve incomplete reconnection \cite{Bateman06}. The present
results are machine independent, so they should apply both to existing
tokamaks and future ones such as ITER.

In future studies, the model should be tested with other extended-MHD
effects such as ion diamagnetic effects and higher $\chi_{||}$. The
restriction on toroidal mode number $n$ should be relaxed. The effect
of the electron pressure profile on the dynamics and the secondary
(drift) instability should be addressed; this may need to utilize
particle-in-cell simulations. Including 3D toroidal geometry is
critical for exploring secondary ideal-MHD instabilities. Comparisons
to multiple tokamak discharges should be done to test the scaling.


We thank I.~T.~Chapman for providing MAST data and thank J.~F.~Drake,
D.~C.~Pace, M.~A.~Shay, and M.~Swisdak 
for helpful conversations.  The authors gratefully acknowledge support
by NSF grant PHY-0902479.  This research used resources at National
Energy Research Scientific Computing Center.


\end{document}